\documentstyle[12pt]{article}
\newcommand{\beq}{\begin{equation}}
\newcommand{\eeq}{\end{equation}}
\newcommand{\iint}{\int_{-\infty}^{+\infty}}
\newcommand{\sech}{\mbox{sech}}
\newcommand{\M}{\mbox{M}}
\newcommand{\sgn}{\mbox{sgn}}
\newcommand{\Hajicek}{H\'{a}j\'\i\v{c}ek}
\newcommand{\Kuchar}{Kucha\v r}
\newcommand{\ReisNord}{Reissner-Nordstr\" om}

\begin{document}
\thispagestyle{empty}
\begin{flushright} 
                   {\small UMDGR--94--76}\\
                   {\small gr-qc/9312033}\\
\end{flushright}

\bigskip

\begin{center}
{\bf \huge Quantum~Gravitational Collapse
of~a~Charged~Dust~Shell}\\[2ex]

\bigskip

Eli Hawkins\footnote{\tt mrmuon@wam.umd.edu}
\medskip

{\small Department of Physics, University
of Maryland, College Park, MD 20742--4111}\\

\bigskip

\end{center}
\begin{abstract}
A simple self gravitating system\,---\,a thin spherical
shell of charged pressureless matter\,---\,is naively
quantized as a test case of quantum gravitational collapse.
The model is interpreted in terms of an inner product on the
positive energy states. An $S$-matrix is constructed
describing scattering between negatively and positively
infinite radius.
\end{abstract}

\section{Introduction}

In the absence of a successful quantum theory of gravity, it
is interesting to study simplified, naively quantized models
of self gravitating systems. Although such a model is not
backed by any complete theory, it may provide clues as to
the properties of a complete theory. Such models may also
confront some of the poorly understood issues of quantum
gravity in a simplified setting.

One interesting gravitational process predicted by General
Relativity is gravitational collapse beyond an event
horizon. A quantitatively accurate model of gravitational
collapse, taking into account the equation of state of the
matter as well as its rotation and other factors is too
complicated to solve in general classically, let alone
attempt to quantize without a full theory. However, some
simple special case of collapse might indicate the
qualitative behavior of the more general case. In this
paper, the equation of state is dealt with by assuming the
collapsing matter to be a pressureless dust. The motion is
simplified by assuming spherical symmetry, but this still
leaves the system with infinite degrees of freedom. If,
instead of a solid ball of matter, the collapsing object is
an infinitesimally thick shell, then it has only one degree
of freedom, its radius, and it is simple enough to deal
with.

Naive quantization of such a dust shell has been
accomplished by \Hajicek, Kay, and \Kuchar\ [2].
Unfortunately, that model is too simplified to exhibit one
of the interesting properties of general gravitational
collapse; in general, if the collapsing object has any
angular momentum or charge it may reexpand, possibly into an
asymptotically flat region distinct from the one in which it
originally collapsed. Allowing the shell to have angular
momentum would break its spherical symmetry and complicate
matters greatly. A charged shell, however, would exhibit
some of the same phenomena, but could still be spherically
symmetric and have only one dynamical degree of freedom.

Unfortunately, a major qualitative difference remains
between the charged and the rotating cases. In the Kerr
space-time there is an accessible, asymptotically flat
region where $r<0$; in the \ReisNord\ metric such a region
exists, but it is only connected to $r>0$ at a single point
of space, and thus hardly seems like a physical part of the
space-time. It seems plausible nonetheless to assume this
spherically symmetric model to be continuous with the more
general case of nonzero angular momentum and therefore to
take the negative radius region seriously.

In this paper I naively quantize the collapse of a charged
dust shell using an equation of motion that is a trivial
generalization of the Wheeler-DeWitt equation derived in
[2]. However, instead of applying a boundary condition at
the origin and only allowing positive radii, I apply a
matching condition at the origin and allow the shell to
attain negative radii.

One issue which this model cannot be expected to resolve is
whether singularities occur in quantum gravity. If there is
a singularity avoidance mechanism, then it may well involve
nonradial degrees of freedom\,---\,even in a spherically
symmetric situation; but all nonradial degrees of freedom
are suppressed in this model.

\section{Equations of Motion}

Consider a spherically symmetric, infinitesimally thick,
dust shell of rest mass $M$ and charge $Q$. The interior of
the shell is a piece of Minkowski space-time; the exterior
is a piece of \ReisNord\ space-time. The mass appearing in
the \ReisNord\ metric is the total energy ($E$\/) of the
system as determined by a distant observer, the charge is
that of the shell. Note that Schwarzschild coordinates on
the interior and exterior do not join smoothly across the
shell ($g_{rr}$ is discontinuous), but this does not cause
any problem in deriving the equation of motion.

The classical equation of motion was derived by \Kuchar\ [1]
from the relationship between the surface energy tensor of
the shell and the difference between the extrinsic
curvatures of its interior and exterior surfaces. The
shell's motion is described by the first integral of motion
\beq
 E = \frac{M}{\sqrt{1-{\dot R}^2}} - \frac{k}{2R}
\label{ceom}
\eeq
where $R$ is the radius of the shell, $\dot R =
\frac{dR}{dT}$, $T$ is the well defined Minkowskian time
coordinate on the interior, and $k = M^2 - Q^2$. Units are
such that \mbox{$\hbar =c=G=1$}. This equation of motion is
formally identical to that of a relativistic point particle
moving radially in a Coulomb potential in flat space-time;
this analogy is useful because it suggests a ready made
interpretation for the quantum model.

By assuming the externally measured energy $E$ to be the
energy conjugate to $T$, numerically equal to the
Hamiltonian that generates motion in $T$, the momentum
conjugate to $R$ is found to be
\beq
P^R = \frac{M \dot{R}}{\sqrt{1 - \dot R^2}}
\eeq
and from this a super-Hamiltonian
\beq
h = -\left(P^T + \frac{k}{2R}\right)^2 + {P^R}^2 + M^2
\eeq
is found which is constrained to vanish and which generates
correct equations of motion. See [2].

Promoting the momenta in $h$ to operators makes the
constraint $h=0$ into a Wheeler-DeWitt equation, namely
\beq
0 = \left(i \frac \partial {\partial T} + \frac k
{2R}\right)^2 \Psi \,
+\, \frac{\partial^2 \Psi}{\partial R^2} - M^2 \Psi .
\label{WdW}
\eeq

Consistent with the above analogy, (\ref{WdW}) is equivalent
to the Klein-Gordon equation for a scalar particle in a
Coulomb potential under the replacement $\Psi =R\Phi$
(equivalent to a change of measure). Equation (\ref{WdW}) is
the basis of this naive quantization.

\section{Probability Interpretation}

This quantum model is most appropriate to an observer
located in the interior of the shell. $T$ and $R$ are well
defined in terms of an observer at $r=0$ who measures the
shell's motion by timing the delays of light pulses
reflected off the inside of the shell. This model may not,
however, be able to describe observations made by an
observer outside the shell. An important effect detectible
only by an external observer would be the shell passing
beyond an event horizon. In this model there is no
qualitative distinction between the motion of the shell
outside and inside an event horizon, as there would be in
a model based upon an exterior observer who only sees
$R>2E$. This is emphasized by
the invariance of both the classical and quantum equations
of motion under the rescaling:
\beq
(T,R,M,E,Q)\longrightarrow
\left (aT,aR,aM,aE,\sqrt{Q^2 - (1-a^2)M^2}\right )
\label{rescale}
\eeq
for any $a>0$. This equates points inside the event horizon
in one case with points outside the event horizon in
another.

Following the scalar particle analogy \Hajicek\ {\em et al}
[2] construct ``charge'' (really number), and energy
currents which are generalized as locally conserved bilinear
currents:
\begin{eqnarray}
J^T(\Psi_1^{},\Psi_2^{}) &=& \frac i2
\left[\bar\Psi^{}_1\dot\Psi^{}_2 -\dot{\bar\Psi}^{}_1
\Psi^{}_2\right] + \frac k{2R} \bar\Psi^{}_1\Psi^{}_2 \\
J^R(\Psi^{}_1,\Psi^{}_2) &=& \frac i2
\left[\bar\Psi^{}_1 \Psi_2^\prime -
\bar\Psi_1^\prime \Psi^{}_2\right] \\
E^T(\Psi^{}_1,\Psi^{}_2) &=& \frac12
\left[\dot{\bar\Psi}^{}_1 \dot\Psi^{}_2 +
\bar\Psi_1^\prime \Psi_2^\prime +
\left(M^2 - \left[\frac k{2R}\right]^2\right)
\bar\Psi^{}_1 \Psi^{}_2\right] \\
E^R(\Psi^{}_1,\Psi^{}_2) &=& \frac {-1}2
\left[\bar\Psi_1^\prime \dot\Psi^{}_2 +
\bar\Psi_1^\prime \dot\Psi^{}_2\right]
\end{eqnarray}
where $^\prime$ denotes $\frac{\partial}{\partial R}$.

Assuming $J(\Psi,\Psi)$ to be a probability current gives a
tentative interpretation of the wavefunction for $R>0$. In
the defining experiment for this model an observer at $r=0$
sends a light pulse at time $U$ which reflects off the shell
at $(T,R)$ and is seen again at time {\em V}.\@ $T$ and $R$
are given by
\beq
\begin{array}{c}
 T=\frac 12 (V+U) \\
 R=\frac 12 (V-U).
\end{array}
\label{UVTR}
\eeq
In the quantum model, if the light pulse is sent at the
observer's time $U$ then the probability that it will return
at time $V$ is $\frac 12 J^UdV$ where $J^U=J^T-J^R$,
evaluated at $(T,R)$ given by (\ref{UVTR}). (I'm assuming an
appropriately normalized wavefunction.)

This interpretation fails, however, because the probability
given is not necessarily positive. Following the scalar
particle analogy further would suggest that the negative
densities might represent pair creation, but this model is
too crude to make a meaningful prediction of that sort. It
would be like modeling a car crashing into a wall as a
potential barrier, and concluding that automobiles would be
pair produced.

\subsection{Boundary Condition}

If only $R\ge 0$ is allowed then $J$ and $E$ will only be
globally conserved if a boundary condition is imposed.
Namely,
 \beq
 J^R(\Psi_1,\Psi_2) = E^R(\Psi_1,\Psi_2)  = 0
\label{bound}
\eeq
at $R=0$ for all $T$.

This boundary condition rules out {\em ab initio} a truly
irreversible collapse. When the shell reaches 0 radius it is
required to pass through itself and reexpand. However, the
collapse may appear irreversible; if the shell reaches $R=0$
it is within an event horizon and an external observer would
never see it reexpand.

Negative radius might seem unphysical, but as I have
mentioned there are plausible reasons for considering it.
Among the stable, approximately spherically symmetric vacuum
space-times (i.e: black holes) the \ReisNord\ space-time is
exceptional. For any such space-time with nonzero angular
momentum the causal structure is that of the Kerr space-
time, and there is an accessible region where $r<0$. In the
quantized case, classical states are smeared and the
\ReisNord\ causal structure might well give in to its
generic neighbor.

In view of the interpretation in terms of an observer at
$r=0$, allowing $R<0$ seems especially meaningless\,---\,an
observer could not possibly continue to exist inside the
shell after it has completely collapsed. However, if the
shell reexpands then observations could be recommenced and
their correlation with previous observations would depend
upon the $R<0$ behavior.

\subsection{Revised Interpretation}

When all values $-\infty<R<+\infty$ are allowed, $J$ gives a
conserved inner product equal to the flux of $J$ across any
Cauchy surface; that is:
\beq
\left<\Psi_1|\Psi_2\right> = \iint J^T(\Psi_1,\Psi_2)\,dR .
\label{product}
\eeq
Because $J^T(\Psi,\Psi)$ is not always positive, it is not
clear that this inner product is positive definite, as is
necessary for normalizable wavefunctions. In fact it is
positive definite if and only if the wavefunctions are
restricted to the subspace of positive energy solutions (see
appendix). This restriction seems acceptable and excuses the
failure of the initial interpretation. The local currents
were not meaningful because the position
operators $R$ or $V$ do not leave the positive energy
subspace invariant and are thus not good observables.

\section{Classical Motion}

First, consider the motion through $R>0$.\@\ In [3] Boulware
analysed the motion of a charged shell through the possible
regions of the extended \ReisNord\ space-time that its
exterior is a piece of. The Schwarzschild coordinates do not
completely specify the position of the shell. The key tool
which helps to distinguish between regions with identical
coordinates is the $r$ component of the unit outward normal
to the outer surface of the shell, which points from the
interior (Minkowski) to the exterior (\ReisNord ) region.
This is given simply by:
\beq
 n^r = \frac EM - \frac{M^2 + Q^2}{2MR} .
\label{nrformula}
\eeq
When $n^r>0$ the exterior of the shell is the $r>R$ region
of \ReisNord\ space-time. When $n^r<0$ the ``exterior'' of
the shell is the $r<R$ region of \ReisNord .

For sufficiently small $R>0$, $n^r$ is necessarily negative.
This means that the shell cannot collapse to form a
singularity of charge $Q$.\@ If it collapses to $R=0$ it in
fact collides with a singularity on its exterior with charge
$-Q$.

Setting $\dot R=0$ in (\ref{ceom}) and solving for $R$ gives
a turning radius
\beq
R_+ = \frac k{2(M-E)} = \frac{M^2 - Q^2}{2(M-E)} .
\eeq

Consideration of $n^r$ and $R_+$ gives the principal
distinct classical trajectories through $R>0$ (along with
trivial cases such as $k=0$).  \\
\indent 1: For $E>M$ and $R_+>0$ (hence ${|Q|}>M$), the
shell starts from $R=\infty$, falls to a finite radius and
bounces back to $\infty$. A further distinction is\\
\indent\indent {\em a}) ${|Q|}>E$ and no event horizon
occurs, or\\
\indent\indent {\em b}) ${|Q|}<E$, an event horizon is formed
and the shell reexpands in a different asymptotically flat
region than it collapsed from.\\
\indent 2: For $E>M$ and $R_+<0$ (hence ${|Q|}<M$), The
shell either falls from $R=\infty$ to 0, or the time
reverse, exploding from 0 to $\infty$.\\
\indent 3: For $E<M$, the shell cannot reach $R=\infty$.
Either,\\
\indent\indent {\em a}) $E<\frac{Q^2+M^2}{2M}$ and the shell
never emerges into an asymptotically flat region, or\\
\indent\indent {\em b}) $E>\frac{Q^2+M^2}{2M}$, the shell
emerges from an event horizon, expands to a maximum radius
of $R_+$, then recollapses, hitting singularities in its
past and future.

Because of the rescaling symmetry (\ref{rescale}), there is
no qualitative distinction between the solutions for the
{\em a} and {\em b} cases. An interior observer can only
guess if the shell has passed within an event horizon.

\subsection{Extension to $R<0$}

Transforming $r \to -r$ in the \ReisNord\ metric shows that
the $r<0$ domain of the complete \ReisNord\ space-time is an
asymptotically flat region containing a singularity with
mass $-E$, and charge either $\pm Q$; by following the
electric field lines through $r=0$ the charge is seen to be
$-Q$. From (\ref{nrformula}) $n^r$ must be positive when
$R<0$; therefore, if the shell enters $R<0$, this
singularity is on the exterior of the shell within a compact
region. The $r$ component of the normal to the inner surface
of the shell also changes sign; the interior thus continues
to be a ball of Minkowski space.

If the classical solution is followed past $R=0$, the
radical in (\ref{ceom}) must change sign; this analytically
continued equation of motion is presumably the only one
continuous with that for nonzero angular momentum. There is
now a second turning point at
\beq
R_- = \frac{-k}{2(M+E)} = -\frac{M^2 - Q^2}{2(M+E)} .
\eeq
For case 2, the shell passes in a finite proper time to
$R=R_-<0$ and then bounces back out to $R=+\infty$. For case
3, The shell oscillates between $R_+>0$ and $R_-<0$ with a
finite period. So, classically, if the shell is allowed to
pass into $R<0$, it will always return to $R>0$ and
potentially be observed again, except in the special case
$k=0$ in which the equation of motion reduces to {$\dot
R=\,$Const.}

\section{Solution of the Wave Equation}

Now find a complete set of solutions to the Wheeler-DeWitt
equation (\ref{WdW}) by considering definite energy states
($P^T$ eigenstates) $\Psi(T,R) = e^{-iET}\psi(R)$. With this
ansatz, (\ref{WdW}) becomes
\beq
0 = \psi^{\prime\prime} +
\left[\left(E+\frac k{2R}\right)^2 - M^2\right] \psi
= \psi^{\prime\prime} + \left[\left(E^2-M^2\right)
+ \frac{Ek}{R} + \frac{k^2}{4R^2}\right] \psi .
\label{deq1}
\eeq

Now assume $E>M$. Defining
\beq
\chi=\sqrt{E^2-M^2}\quad ;\quad x=\chi R
\eeq
(\ref{deq1}) becomes
\beq
0 = \frac{d^2\psi}{dx^2} + \left[1 +
\frac{Ek}{\chi x} + \frac{k^2}{4x^2}\right] \psi .
\label{deq2}
\eeq
The characteristic equation about the regular singularity at
$x=0$ is
\beq
0 = \lambda(\lambda-1) + \frac{k^2}4
\eeq
with roots $\lambda_\pm = \frac{1\pm\alpha}2$ where
$\alpha=\sqrt{1-k^2}$. As noted in [2], this model breaks
down for ${|k|}>1$, corresponding to an overcritical Coulomb
potential which can only be described by a second quantized
theory. It is far from clear what an appropriate analogue of
second quantization would be here, and it also seems
implausible that an analysis of higher order effects could
be meaningful in a model with virtually all the system's
degrees of freedom suppressed. I therefore only consider
${|k|}<1$.

First, only consider the solution of (\ref{deq2}) in $x>0$;
there must exist independent solutions such that
\beq
\psi = x^{\lambda_\pm} \left(1+O(x)\right) .
\label{about0}
\eeq
Because the coefficients in (\ref{deq2}) are real, the
solutions (\ref{about0}) are real for all $x>0$. Adopting
one more new constant $\beta=\frac{Ek}{2\chi}$, (\ref{deq2})
may be written
\beq
0 = \frac{d^2\psi}{d(2ix)^2} + \left[-\frac14
- \frac{i\beta}{2ix} + \frac{\frac14 -
\frac14\alpha^2 }{(2ix)^2}\right] \psi
\label{deq3}
\eeq
which is just Whittaker's equation (see [4] p505). Define
two functions, equal to the solutions (\ref{about0}):
\beq
\varphi_\pm(\alpha,\beta;{x>0}) =
e^{\pm ix} x^{\lambda_\pm}
\M\left({\lambda_\pm+i\beta},{2\lambda_\pm},{2ix}\right)
\eeq
where $\M$ is the regular Kummer confluent hypergeometric
function. These solutions are indeed real (see [4] eq.
13.1.27).

$\M$ is a single valued function, but because of the
$x^{\lambda_\pm}$ factor $\varphi_\pm$ are generally multi-%
valued functions in the complex plane with a branch point at
$x=0$. In order to analyse the $R<0$ behavior it must be
determined which branch the physical $-x$ axis lies on.
(This analysis also applies when $E<M$.) As a criterion,
note that (\ref{WdW}) is invariant under the time reversal
transformation $\Psi(T,R) \to \bar\Psi(-T,R)$ or
equivalently $\psi(R) \to \bar\psi(R), E\to E$. Because of
this we will demand that the complex conjugate of any
physical solution of (\ref{deq1}) must also be a physical
solution. This means that $\varphi_\pm$ must be real for
$x<0$ as well as $x>0$, otherwise $\varphi_\pm$ and
$\bar\varphi_\pm$ would constitute more than 2 independent
solutions to (\ref{deq1}).

A second criterion is provided by the current form $J^R$.
For two definite energy states of the same energy,
conservation implies that $J^R$ must be constant with
respect to $R$. So
\beq
\varphi_+(\alpha,\beta;x) \varphi^\prime_-(\alpha,\beta;x) -
\varphi^\prime_+(\alpha,\beta;x) \varphi_-(\alpha,\beta;x)
= \mbox{Const.}(x) .
\eeq
This implies that one of the $\varphi_\pm$ is approximately
even, and the other approximately odd near $x=0$; it doesn't
matter which. Arbitrarily choose $\varphi_+$ as
approximately even\footnote{Strictly speaking, this sign
combination (or its alternative) cannot be attained as a
branch of the multi-valued solution for certain rational
values of $\alpha$, but the complement of that set of values
is dense, and this is physics.}.

Now note that (\ref{deq3}) is invariant under
$(x,\beta)\to(-x,-\beta)$ so
\beq
\varphi_\pm(\alpha,\beta;x)=
\pm\varphi_\pm(\alpha,-\beta;-x) .
\label{flip}
\eeq

\section{Transmission and Reflection coefficients}

In a definite energy state for $R\to\pm\infty$,
$J^T(\Psi,\Psi)$ becomes simply $E\cdot{|\Psi|}^2$; the
interpretation simplifies. For a fixed energy, the
wavefunction near infinity may be interpreted as a relative
probability amplitude, and an $S$-matrix may be constructed.
Time reversal invariance and unitarity restrict the $S$-
matrix to the form:
\beq
S = \left(\begin{array}{cc} S_{++}& S_{+-}\\
S_{+-}&S_{++}  \end{array} \right)
\eeq
where
\beq
\begin{array}{ccc}
S_{++} &=& e^{i\delta} \cos \gamma\\
S_{+-} &=& i e^{i\delta} \sin \gamma .
\end{array}
\eeq

If (\ref{deq1}) is viewed as a time independent
Schr\"odinger equation, the effective potential is long
ranged, and so the wavefunctions cannot asymptote to
definite momentum waves with stable phase. As a result the
phase $\delta$ is not well defined; only the reflection and
transmission coefficients $\cos^2\gamma$ and $\sin^2\gamma$
are meaningful. The reflection coefficient is interpreted as
the probability that a shell collapsing from $+\infty$ will
return to $+\infty$; assuming this model to have any
validity, that probability should be independent of how the
position of the shell is measured.

These reflection and transmission coefficients for the
scattering of wave packets between $+\infty$ and $-\infty$
are determined from the asymptotic behavior of the two basis
solutions $\psi(R) = \varphi_\pm(\alpha,\beta;\chi R)$.

Equation 13.5.1 of [4] gives the asymptotic behavior of
$\varphi_\pm$:
\beq
\varphi_\pm(\alpha,\beta;x\to+\infty)=
{(\pm\alpha)!}\left[{\left(\frac i2\right)}^{\lambda_\pm}
\frac{{e^{- \frac\pi{2} \beta}} {e^{-ix}} {(2x)^{-i\beta}}}
{\Gamma{\left(\lambda_\pm - i\beta\right)}} + {c.c.}\right]
\left(1+O\left(x^{-1}\right)\right)
\eeq
and by (\ref{flip}),
\beq
\varphi_\pm(\alpha,\beta;x\to-\infty)=
\pm{(\pm\alpha)!}\left[{\left(\frac i2\right)}^{\lambda_\pm}
\frac{{e^{ \frac\pi{2} \beta}} {e^{ix}} {(-2x)^{-i\beta}}}
{\Gamma{\left(\lambda_\pm + i\beta\right)}} + {c.c.}\right]
\left(1+O\left(x^{-1}\right)\right)
\eeq

In lieu of definite momentum waves, the appropriate ingoing
and outgoing asymptotic waves are $e^{\pm i \phi}$ where
\beq
\phi = \chi R + \sgn(R)\,\beta \ln{|2\chi R|}
\eeq

In terms of $\phi$, the asymptotic forms of the
wavefunctions are
\beq
\psi(R\to+\infty) ={(\pm\alpha)!}\left[{\left(\frac i2
\right)}^{\lambda_\pm} \frac{{e^{- \frac\pi{2} \beta}} }
{\Gamma{\left(\lambda_\pm - i\beta\right)}}
{e^{-i\phi}} + {c.c.}\right]
\left(1+O\left(R^{-1}\right)\right)
\eeq
\beq
\psi(R\to-\infty) =\pm{(\pm\alpha)!}\left[{\left(\frac i2
\right)}^{\lambda_\pm} \frac{{e^{ \frac\pi{2} \beta}}}
{\Gamma{\left(\lambda_\pm + i\beta\right)}}
{e^{i\phi}} + {c.c.}\right]
\left(1+O\left(R^{-1}\right)\right)
\eeq
which are of the form
\beq
\begin{array}{c}
\psi(r\to+\infty)\longrightarrow
c_{\mbox{\scriptsize in}-} e^{i\phi} +
c_{\mbox{\scriptsize out}+} e^{-i\phi}\\
\psi(r\to-\infty)\longrightarrow
c_{\mbox{\scriptsize in}+} e^{-i\phi} +
c_{\mbox{\scriptsize out}-} e^{i\phi}
\end{array}
\eeq
By applying the definition of the $S$-matrix:
\beq
\left(\begin{array}{c}c_{\mbox{\scriptsize out}+}\\
c_{\mbox{\scriptsize out}-}\end{array}\right)
= S \left(\begin{array}{c}c_{\mbox{\scriptsize in}+}\\
c_{\mbox{\scriptsize in}-}\end{array}\right)
\eeq
one obtains,
\beq
i e^{\pm\frac \pi2 i \alpha} = \pm S_{++} e^{\pi\beta} +
 S_{+-} \frac{\Gamma\left(\lambda_\pm - i\beta\right)}
{ \Gamma\left(\lambda_\pm + i\beta\right)}
\eeq

Combining the $+$ and $-$ equations and solving gives
\beq
\delta = \arg\left[\Gamma\left(\lambda_\pm - i\beta\right)
\Gamma\left(\lambda_\pm + i\beta\right) \right]
\eeq
\beq
\cos^2\gamma = \sin^2
\left(\frac\pi2\alpha\right) \sech^2{\pi\beta} .
\eeq

Classically, the shell cannot reach $R=-\infty$, unless
$k=0$ in which case it must. Correspondingly, the
transmission coefficient, $\cos^2 \gamma$, is necessarily
$<1$ unless $k=0$, in which case it is necessarily $=1$. The
transmission coefficient goes to 0 when $k\to\pm1$ or $E\to
M$.

There are poles in the transmission coefficient when $\cosh
\pi\beta=0$; accepting only normalizable (positive energy)
solutions, this implies the existence of bound states with
energies
\beq
E_n = \frac{M (2n-1)} {\sqrt{(2n-1)^2 + k^2}}
\eeq
with $n=1,2,3,\ldots$.\@ This is a different energy spectrum
than results if the boundary condition (\ref{bound}) is
imposed. For $k<0$, the bound states correspond to states
with no asymptotically flat region. For $k>0$ the bound
states all correspond to classical states with maximum
radius ($R_+$) outside the event horizon (classical case
3{\em b}). Because this model is unable to deal with an
observer outside the shell, it cannot decide questions of
when the shell does or does not pass beyond an event horizon
any better than this.

The bound state solutions can be found explicitly in the
same way as the scattering solutions, but in contrast to the
bound states found in [2] with the boundary condition
(\ref{bound}) imposed, they are not algebraically special.

\section{Conclusions}

By necessity, this Quantum model is based upon several
tentative assumptions. The standard Schwarzschild time $t$
is the proper time of a distant, stationary observer; such
an observer will measure the total mass of the system to be
$E$; therefore $E$ is the energy conjugate to $t$. The first
assumption made in this model was that $E$ is conjugate to
$T$. This works if there is a time slicing such
that $T$ and $t$ differ only by a constant; that may be
accomplished locally but not necessarily globally in this
system. This issue will be further addressed for the
uncharged case in [5]. If the correct energy for $T$ is not
$E$, then it must be some constant of motion invariant under
time translations (as $P^T$ commutes with itself); this
restricts it to be some homogeneous function of $E$, $M$,
and $Q$.

The second assumption is that a Wheeler-DeWitt equation
based on the super-Hamiltonian is the correct method of
quantization; this is probably not completely true and is
the naive aspect of this model.

The third assumption is that, based on the equivalence of
the equations of motion with those of a scalar particle, the
inner product (\ref{product}) gives the correct probability
interpretation of the wavefunctions.

The restriction of my interpretation to measurements made by
an observer on the interior is necessary. One might attempt
to apply this model to observations made from $\infty$; if a
distant observer reflects light off of the shell (to be
precise a pulse is sent from some ${\cal I}^-$ and received
at some ${\cal I}^+$, possibly in a different asymptotically
flat region), then the appropriate coordinates are the
Kruskal coordinates of the shell in the exterior region.
Unfortunately, the transformation to these coordinates from
$T$ and $R$ depends upon the whole trajectory of the shell;
there is no obvious quantum equivalent. A model analogous to
this one, but based upon an outside observer appears
intractable. The first step to constructing it would be to
write $E$ as a function of $R$ and $\frac{dR}{dt}$, but this
is the solution of a quartic equation and is rather
unwieldy.

This restriction to interior observers means that this model
cannot directly deal with the interesting questions of when
(or rather with what amplitude) the shell will pass beyond
an event horizon and remerge in a different asymptotically
flat region. This can only be addressed crudely by comparing
quantum states with classical trajectories. A typical
wavepacket will be a combination of definite energy states
whose classical equivalents do and do not contain event
horizons. This combination can be crudely interpreted to
give the probability of an event horizon occurring.

This model has in some sense addressed the issue of whether
singularities occur. When $R<0$, the corresponding classical
space-time will include a singularity at $r=0$. However, the
dynamics of this model are only mildly singular at $R=0$,
mildly enough that the solution could be simply continued
across that singularity. This might be taken as a hopeful
sign that the dynamics of a full quantum theory of
gravitation will be singularity free.

\section{Appendix}

Proposition: {\em The restriction to positive energies makes
the inner product (\ref{product}) positive definite}.

To see this consider a state $\Psi=e^{-iET}\psi(R)$ with
definite energy $E>0$ (it is straightforward to show that
different energy states are orthogonal). We have
\beq
\left< \Psi | \Psi \right> = \iint
\left(E+\frac k{2R}\right) {|\psi|}^2{dR}.
\eeq
For scattering states this must be positive as
${|R|}\approx\infty$ contributions dominate. For bound
states $\psi(R\to\pm\infty)\longrightarrow0$, hence for $a>0$,
\begin{eqnarray*}
0&=&\left[(\bar\psi(aR))^\prime \psi(R) -
\bar\psi(aR) \psi^\prime(R)\right]_{-\infty}^{+\infty}\\
&=&\iint\left[(\bar\psi(aR))^{\prime\prime} \psi(R) -
\bar\psi(aR) \psi^{\prime\prime}(R)\right]dR\\
&=&\iint\left[\left(E+\frac k{2R}\right)^2 -
\left(aE+\frac k{2R}\right)^2 -
\left(1-a^2\right) M^2\right] \bar\psi(aR)\psi(R) \, dR\\
&=&(1-a)\iint \left[E\left([1+a]E + \frac kR \right) -
(1+a) M^2\right] \bar\psi(aR)\psi(R) \, dR,
\end{eqnarray*}
so for $a\ne1$
\beq
0 = \iint\left[E\left([1+a]E + \frac kR\right) -
(1+a)M^2\right] \bar\psi(aR)\psi(R) dR.
\eeq
This integrand diverges at $R\to0$ as $R^{-\alpha}$ (recall
${|\alpha|}<1$) and approaches 0 exponentially as
$R\to\pm\infty$. As a result, this improper integral may be
uniformly approximated to arbitrary accuracy by a proper
integral and is therefore a continuous function of $a$ and
so
\beq
\left<\Psi | \Psi\right> =
\frac{M^2}E \iint {|\psi(R)|}^2 dR>0
\eeq
which is manifestly positive. Similarly this is clearly
negative for $E<0$; hence restriction to positive energy is
both necessary and sufficient for the inner product to be
positive definite.

\section*{Acknowledgements}

I wish to thank T. Jacobson, K. \Kuchar\, and P. \Hajicek\
for invaluable discussions, and J. Friedman for suggesting
this topic. In addition, I thank the Institute for
Theoretical Physics at Santa Barbara where most of this
work was done.

Research supported by NSF REU award under grant PHY91-12240
and at the ITP under grant PHY89-04035.

\section*{References}
{[1]} K. \Kuchar\, Czech. J. Phys. B {\bf 18}, 435 (1968)\\
\\
{[2]} P. \Hajicek\, B. S. Kay, K. V. \Kuchar\, Phys. Rev. D
{\bf 46}, 5439 (1992)\\
\\
{[3]} D. G. Boulware, Phys. Rev. D {\bf 8}, 2363 (1973)\\
\\
{[4]} M. Abramowitz and I. Stegun, {\em Handbook of
Mathematical Functions} (Washington :  U.S. Govt. Print.
Off.,  1972)\\
\\
{[5]} P. \Hajicek\, to be published.

\end{document}